\documentclass{INTERSPEECH2023}
\usepackage{subfigure}
\usepackage{amsmath,graphicx,epsfig,stfloats,hyperref,multirow,booktabs,caption,color,cite,amssymb}
\usepackage{amsfonts}
\usepackage{marvosym}
\interspeechcameraready

\title{CQNV: A combination of coarsely quantized bitstream and neural vocoder for low rate speech coding}
\name{Youqiang Zheng$^1$, Li Xiao$^{1,\dagger}$,Weiping  Tu$^{1,2,3}$\textsuperscript{,\Letter}, Yuhong Yang$^{1,3}$, Xinmeng Xu$^{1}$ \thanks{\Letter Corresponding author: Weiping Tu(tuweiping@whu.edu.cn)}
\thanks{$\dagger$ Equal contribution. The numerical calculations in this paper have been done on the supercomputing system in the Supercomputing Center of Wuhan University.}}

\address{
  $^1$NERCMS,
School of Computer Science, Wuhan University, China\\
  $^2$Hubei Luojia Laboratory,  China\\
  $^3$Hubei Key Laboratory of Multimedia and Network Communication Engineering, \\Wuhan University, China
  }
  
\email{\{youqiangzheng, xiaoli1996, tuweiping, yangyuhong, xuxinmeng\}@whu.edu.cn}

\begin{document}

\maketitle

 
\begin{abstract}
Recently, speech codecs based on neural networks have proven to perform better than traditional methods. However, redundancy in traditional parameter quantization is visible within the codec architecture of combining the traditional codec with the neural vocoder. In this paper, we propose a novel framework named CQNV, which combines the coarsely quantized parameters of a traditional parametric codec to reduce the bitrate with a neural vocoder to improve the quality of the decoded speech. Furthermore, we introduce a parameters processing module into the neural vocoder to enhance the application of the bitstream of traditional speech coding parameters to the neural vocoder, further improving the reconstructed speech's quality. In the experiments, both subjective and objective evaluations demonstrate the effectiveness of the proposed CQNV framework. Specifically, our proposed method can achieve higher quality reconstructed speech at 1.1 kbps than Lyra and Encodec at 3 kbps.
\end{abstract}

\section{Introduction}
Speech coding has emerged as a prevalent technique in various communication fields, such as mobile and satellite communication~\cite{McCree2008}. Its ability to significantly diminish the bandwidth and storage costs needed for information transmission and storage is well-known~\cite{noll1997mpeg}. Existing traditional parametric codecs based on the linear prediction model have achieved significant success at bitrates above 2.4 kbps, such as MELP~\cite{supplee1997melp} and Codec2~\cite{rowe2011codec2}. These codecs rely on codebooks to quantize speech parameters and generate speech perceptually similar to the original~\cite{o2023review}. However, as the bitrate is reduced, traditional parametric codecs face a performance challenge due to the inevitable increase in quantization error.

With advanced deep learning, the deep neural network has been universally applied to speech coding tasks, leading to promising results. In particular, the end-to-end architecture employs an encoder-decoder architecture, where both the encoder and decoder utilize neural networks and are jointly trained. This method has achieved a high quality while passing speech through a compact latent representation corresponding to very low bitrates ~\cite{garbacea2019low,2021soundstream,2022nesc,TFNet,2022encodec}. Among them, Soundstream~\cite{2021soundstream} is a whole end-to-end approach that compresses 24 kHz audio at the bitrates from 3  kbps to 18 kbps. Encodec~\cite {2022encodec}, on the other hand, employs a streaming encoder-decoder architecture with quantized latent space and introduces a Transformer~\cite{vaswani2017attention} language model to further compress the obtained representation by up to 40\%. It reduces the bandwidth of the 1.5 kbps model to 0.9 kbps and demonstrates superior performance compared to Opus~\cite{valin2012opus} at 6 kbps. While end-to-end techniques have proven superior to conventional methods in low-bitrate speech coding, their black-box nature poses a significant challenge for model interpretation and improvement. Therefore, analyzing and enhancing end-to-end models is still a daunting task.

The neural vocoder-based approach, on the other hand, combines traditional and deep learning-based methods and provides more interpretability compared to the end-to-end architecture. Typically, the bitstream is generated by a traditional encoder and decoded by a neural vocoder. In~\cite{kleijn2018wavenet},  WaveNet~\cite{oord2016wavenet} is the first explored as a generic generative model combined with Codec2 for speech coding, which achieves high performance at 2.4 kbps. Similarly, vocoders, including SampleRNN~\cite{2016samplernn}, and LPCNet~\cite{2019lpcnet} have been used in speech coding with a traditional encoder and present good performance \cite{2019samplernn,lpcnet-1.6kbps}.
In order to boost speed in synthesizing speech, StyleMelgan~\cite{2021stylemelgan} has been used to decode the bitstream of LPCNet at 1.6 kbps and generated state-of-the-art quality with low delay~\cite{SSMGAN} by a high degree of parallelization. 

The neural vocoder-based approaches mentioned above have demonstrated that in combination with parametric codecs, neural vocoders can achieve good quality in low-bitrate speech coding. However, we summarise two main drawbacks based on these methods: 1) \textbf{Redundancy in parameter quantization.} The bitrates associated with existing research \cite{kleijn2018wavenet} at 2.4 kbps and \cite{2019lpcnet} at 1.6 kbps, exceed 1.2 kbps compared to the end-to-end approach, which implies the bitstream of traditional speech coding parameters exists redundancy that is not normally used by the neural vocoder. 2) \textbf{Possible performance limitations.} The neural vocoder employs a frame-based intermediate acoustic representation as its input, such as mel-spectrograms. However, the neural vocoder used in speech coding primarily relies on traditional speech coding parameters as input. This divergence in application contexts may impose limitations on the performance of the vocoders. Thus, a direct combination of bitstream from parametric codecs with a neural network vocoder may not be optimal.

The analysis above motivates us to design a new framework that decreases the bitrate while significantly enhancing the quality of the decoded speech. In this paper, we propose a framework that involves coarsely quantizing the speech parameters obtained from Codec2 to reduce the bitrate. We explore three ways of coarse quantization with different parameters. The coarsely quantized parameters are then fed into the HiFi-GAN~\cite{hifigan} vocoder at a minimum of 1 kbps, which can solve the first drawback. This framework simplifies the codebook design of quantizers in traditional parametric codecs and achieves high-efficiency coding. Meanwhile, a parameters processing module is presented to tackle the second drawback by adopting
three branches to carry different dilation rates of convolutional
layers, which can achieve high-quality decoded speech reconstruction. This approach can significantly overcome the drawbacks mentioned above. The contributions of this study are summarized as follows:
\begin{itemize}
\item We explore the ability of the neural network-based vocoder to produce speech with coarsely quantized parameters of the traditional speech codecs, which provides a simplified approach to reduce the bitrate of speech while maintaining high quality.
\item  We present a parameters processing module that enhances neural vocoders' quality based on the bitstream of traditional speech coding parameters. Meanwhile, we introduce a dynamic hyper-parameter to balance the losses during training. 
\item Experiments show that our method is preferred to previous approaches even if they use more than 3x the bitrate.
\end{itemize}

\section{Method}
\begin{figure*}[ht]
\centering
\subfigure[]{\includegraphics[width=0.305\textwidth]{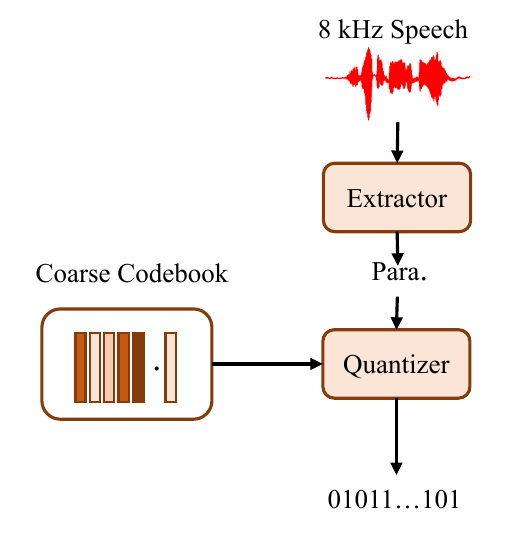}}
\subfigure[]{\includegraphics[width=0.50\textwidth]{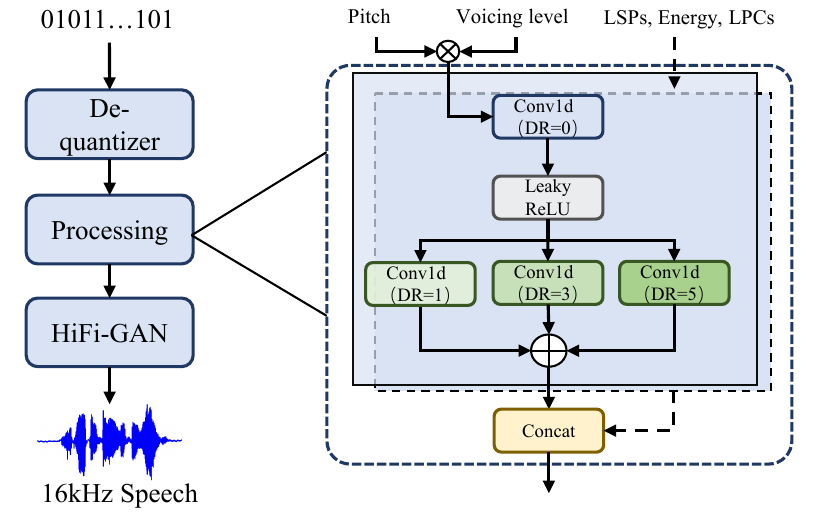}}
\caption{The overall architecture of the proposed models. Parameters are coarsely quantized to a bitstream at (a) the encoder. At (b) the decoder bitstream is de-quantized to parameters, and it passes through the processing module as the conditional input of the vocoder.  }
\label{fig:structure}
\end{figure*}

As shown in Figure \ref{fig:structure},
the proposed framework consists of Codec2 components (encoder and de-quantizer) and HiFi-GAN vocoder. The parameters quantizer is redesigned and trained on the encoder, which uses fewer bits to quantize the speech parameters extracted from the source speech sampled at 8 kHz. On the decoder,  HiFi-GAN synthesizes 16 kHz speech with the parameters from the de-quantizer.
\subsection{Encoder}
Codec2 extracts parameters from each frame of the speech with a length of 10 ms. When working at 1.2 kbps, the parameters of every four consecutive frames are jointly quantized. Furthermore, 27, 16, and 4 bits are used to quantize the following parameters:  line spectrum pairs (LSPs), pitch and energy, and voicing level. In order to reduce bitrate and explore the generative power of the neural vocoder with less input information, we propose a coarse quantization of LSPs, pitch and energy. For the voicing level, each frame is characterized by one-hot encoding, 
so it is not possible to cut the bitrate further.

\noindent

\textbf{\emph{The quantization of LSPs}} :
LSPs are sampled every four frames, and the three frames in between are interpolated at the de-quantizer. We use two-stage vector quantization \cite{gray1984vector} and split vector quantization for LSP, which keeps the same with Codec2 1.2 kbps. The 10-order LSPs are quantized in the first stage by using a codebook with the size of 512, which indicates that 9 bits are used to index the code vectors. And the quantization residual is computed. In the second stage, the quantization residual is split into two 5-dimension vectors, composed respectively of the odd-order and even-order components of the residual vector, and then quantized independently using two codebooks with the same size of 128 rather than 512. This operation indicates that index bits decrease from 9 to 7. All 
 three codebooks are trained by using the LBG algorithm\cite{lbg1980}. The quantization of LSPs achieves a reduction in the number of bits from 27 to 23 for four consecutive frames.

\noindent
\textbf{\emph{Joint quantization of Pitch and Energy}} : 
Pitch and energy are quantized by joint vector quantization. They are sampled every two frames, and the frame in between is interpolated in the decoder. The dimension of the codebook is 2. The first item is the $\log_{2}$ of the pitch compared to the lowest frequency, and the second item is the energy in dB, defined in Equation \ref{Wo} and Equation \ref{E}, where \textbf{Wo} is the pitch, and \textbf{e} is the energy. 
\begin{equation}
x_p={\log _{2}\left(\frac{Wo}{\pi} \cdot \frac{4000}{50}\right)}
\label{Wo}
\end{equation}
\begin{equation}
x_e=10 \times \log _{10}\left(e+10^{-4}\right)
\label{E}
\end{equation}

A codebook with a size of 256 is used in Codec2 at 1.2 kbps for quantizing pitch and energy. We designed and trained a new codebook with a size of 64, reducing the bits from 16 to 12 for a packet composed of 4 continuous frames. The codebook is designed using a method that combines prediction and vector quantization\cite{jmv-codec2}, which uses the prediction residuals and inter-frame correlation to improve the performance of the codebook, i.e., the prediction vector of the current frame is based on the prediction coefficients and the previous frame. The prediction coefficients are 0.8 for pitch and 0.9 for energy, following Codec2. Our codebook is trained on the residual vectors calculated by prediction and initial vectors. Moreover, weighted errors are computed for different features, e.g., pitch and energy errors are given much higher weights on stationary speech than on non-stationary speech or silence.
\subsection{Decoder}

In the decoder, a de-quantizer obtains parameters from the bitstream. Then the parameters are introduced to condition the HiFi-GAN model at 100 Hz to produce speech signals. To condition HiFi-GAN, this work uses 23-dimension features, LSPs, energy, pitch, voicing level, and linear prediction coefficients (LPCs) calculated from LSPs. Instead of providing the features straight to the network, we employ a module as shown in Figure \ref{fig:structure}(b) to process the parameters. In this module,
we adopt three branches to carry different dilation rates of convolutional layers to generate feature maps  with different receptive field sizes. More precisely, the pitch is multiplied elementwise by voicing level to obtain a representation through this module. It is processed separately from the rest of the conditioning parameters because the pitch is critical for high-quality synthesis. Then LSPs, energy, and normalized LPCs parameters are passed through the same block to obtain another representation. Two representations are passed through concatenation to condition the HiFi-GAN. With the module, the neural vocoder based on the bitstream of traditional speech coding parameters can effectively reconstruct speech.

HiFi-GAN, a generative adversarial network consisting of one generator and two discriminators, achieves more efficient and high-fidelity speech than auto-regressive or flow-based models. It is trained to produce an audio waveform when conditioned on mel-spectrograms, along with GAN losses and two additional feature losses for improving training stability and model performance. Following the similar research \cite{yang2021ganspeech}, we introduce a dynamic ratio of the feature matching loss $\lambda_{\mathrm{FM}}$ to replace the fixed hyper-parameter used in \cite{hifigan}. It is the ratio of the mel-spectrograms loss to the feature matching loss during training. Our final objectives for the generator and discriminator are as follows:
\begin{equation}
\mathcal{L}_G=\mathcal{L}_{A d v}(G ; D)+\lambda_{f m} \mathcal{L}_{F M}(G ; D)+\lambda_{m e l} \mathcal{L}_{M e l}(G)
\end{equation}
\begin{equation}
\mathcal{L}_D=\mathcal{L}_{A d v}(D ; G)
\end{equation}
, where we set $\lambda_{m e l}= 45$ and $\lambda_{f m} = \mathcal{L}_{M e l}(G) / \mathcal{L}_{F M}(G ; D) $. This implementation effectively balances the losses and improves convergence speed during training.

\section{Experimental Setup}
\begin{figure*}[ht]
\centering

\subfigure[]{\includegraphics[width=0.21\textwidth]{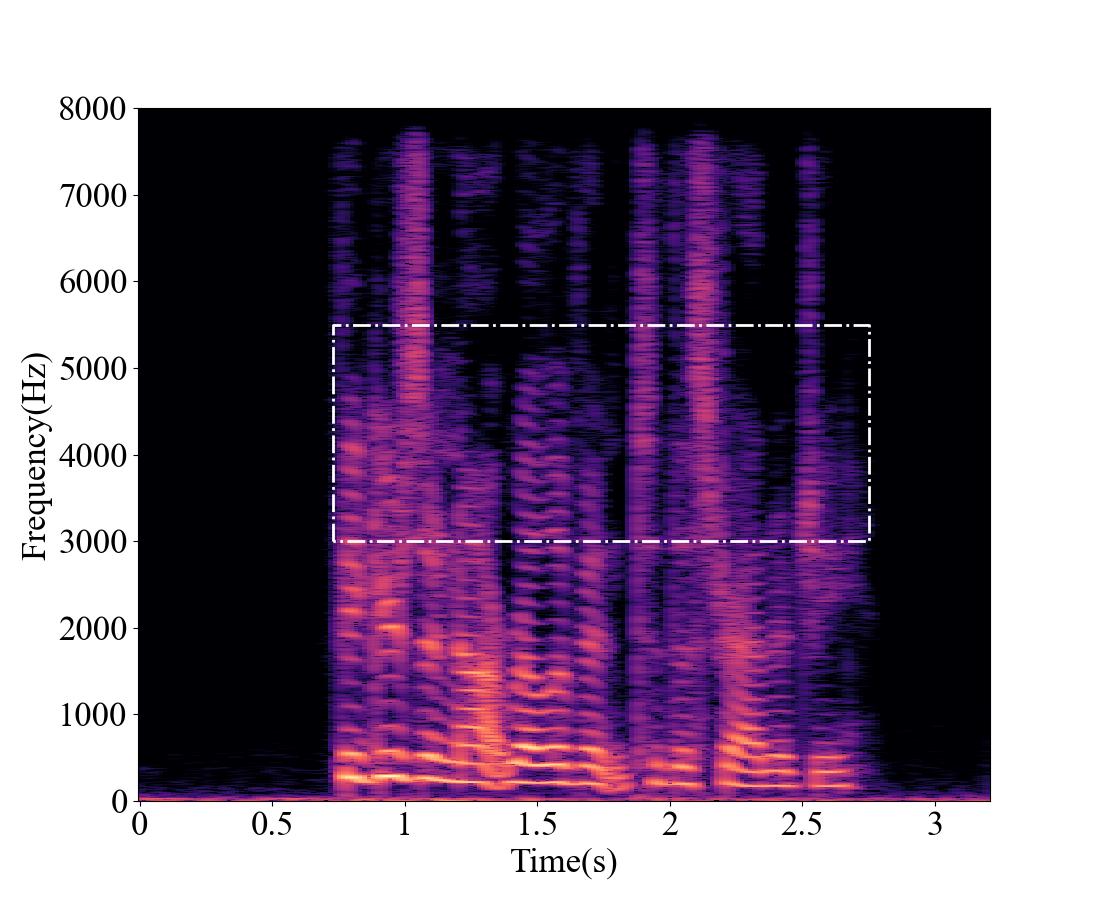}}
\subfigure[]{\includegraphics[width=0.21\textwidth]{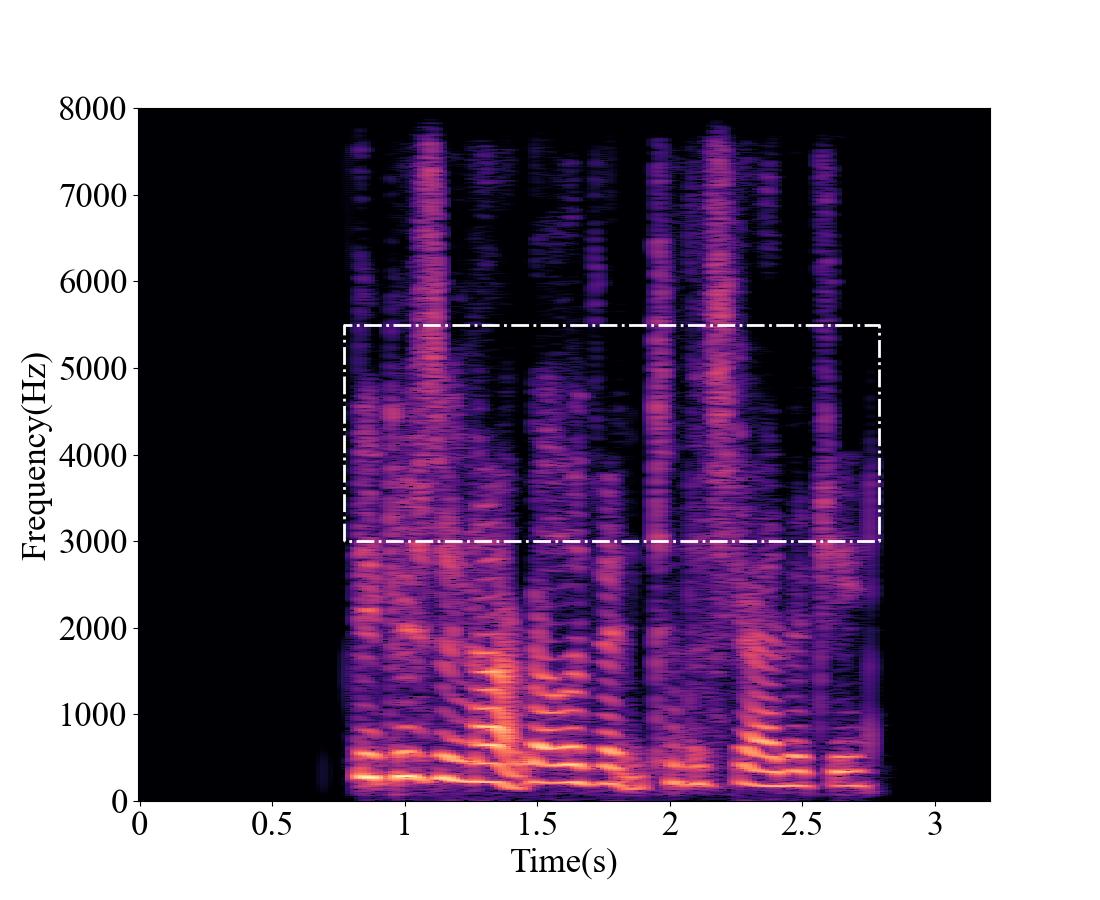}}
\subfigure[]{\includegraphics[width=0.21\textwidth]{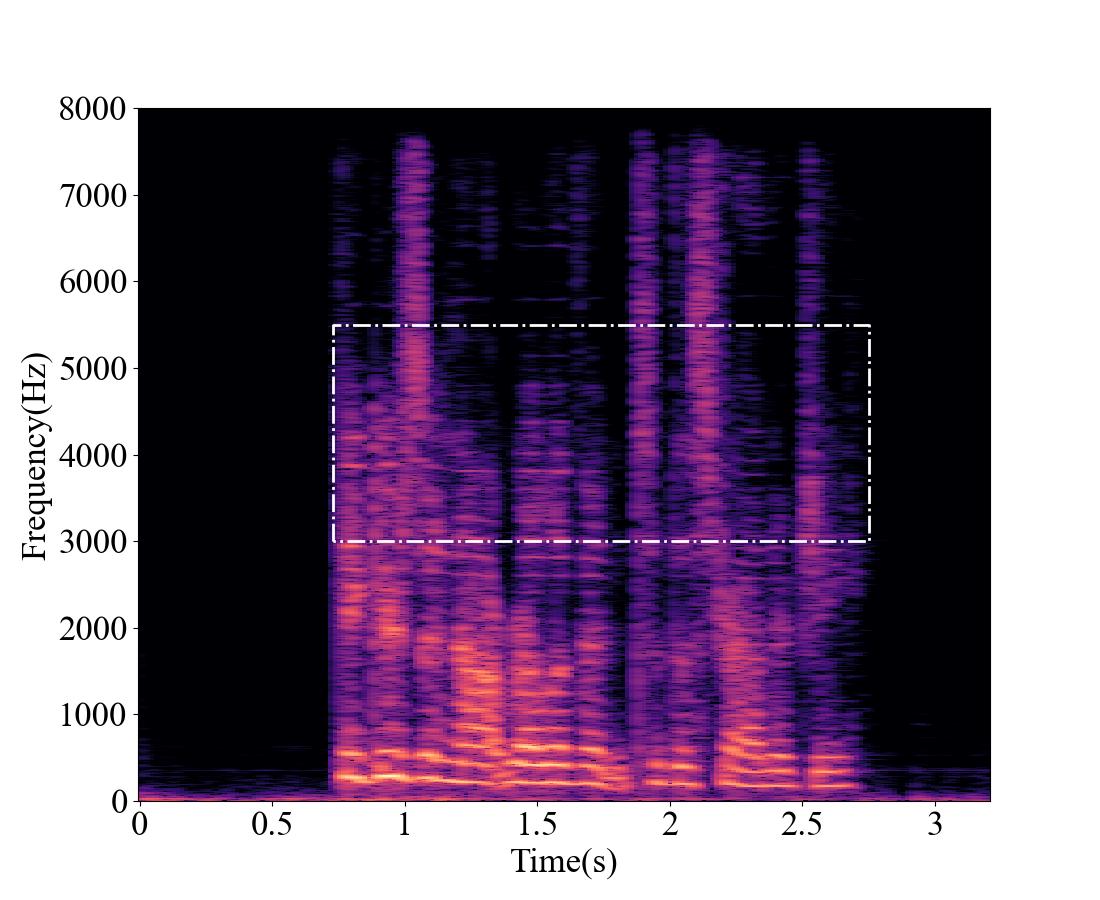}}
\subfigure[]{\includegraphics[width=0.21\textwidth]{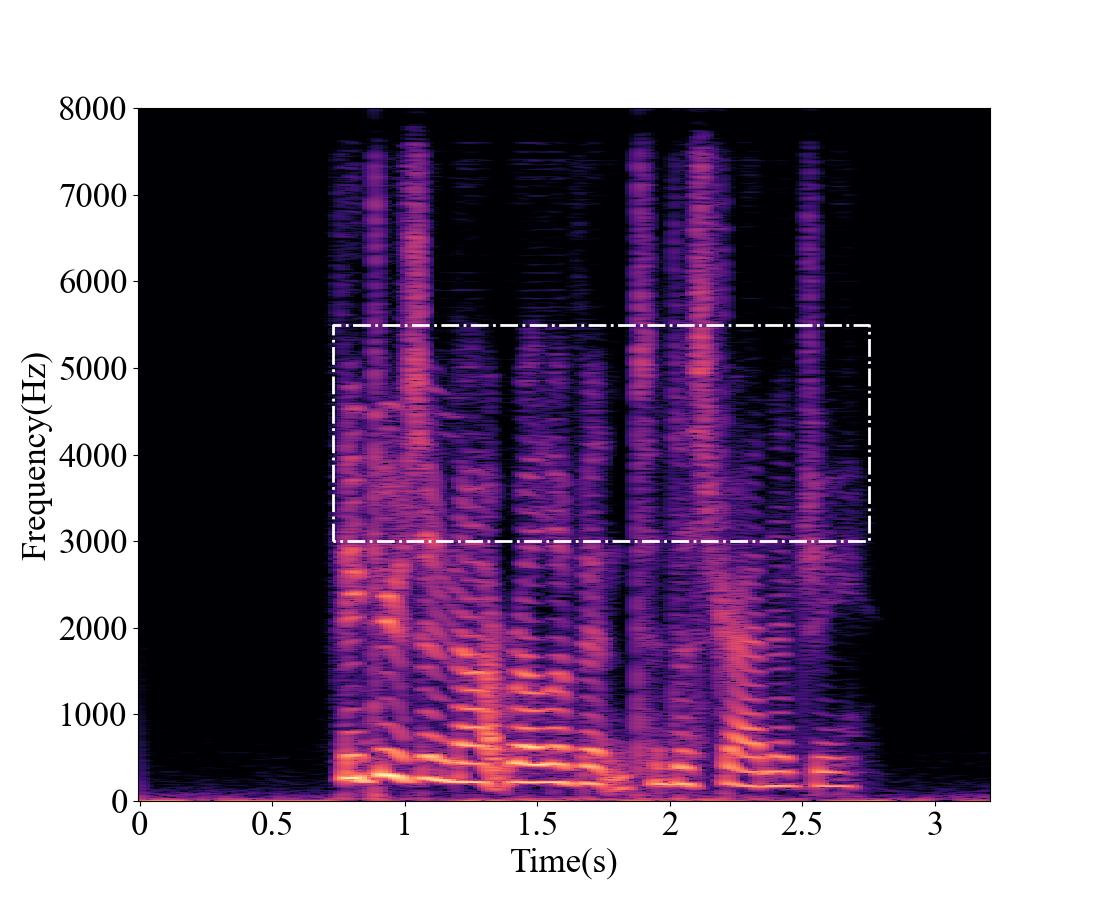}}

 \caption{ Comparing spectrograms of a test sample. Starting from (a) 16 kHz ground truth, (b) Lyra at 3 kbps (c) Encodec at 3 kbps, (d) CQNV-v3 at 1  kbps.}
 \label{fig:yuputu}
\end{figure*}
\subsection{Datasets}

We use VCTK multi-speaker datasets~\cite{VCTK} to train and evaluate the proposed coding method. The total length of the audio clips is approximately 40 hours, and the sample rate of the audio is 44.1 kHz. We downsample the speech data to 8 kHz for training codebook and 16 kHz for training HiFi-GAN. The codebook and HiFi-GAN are trained with the same corpus from 100 speakers, including 57 female and 43 male speakers, and tested with 8 unseen speakers, including 4 females and 4 males.

\subsection{Implementation}

Four codebooks are trained, three for quantizing LSPs and one for pitch and energy parameters. Based on the coding architecture of Codec2 at 1.2 kbps and the newly trained codebooks, three versions of the coding scheme with coarsely quantized parameters are developed. Table.\ref{tab:bit allocation} shows the bit allocation of parameters in different versions for each packet with a length of 40 ms. \textbf{i)} \textit{CQNV-v1}: bits of LSPs are reduced from 27 to 23, and the bitrate is reduced to 1.1  kbps; \textbf{ii)} \textit{CQNV-v2}: bits of pitch and energy is reduced from 16 to 12, and the bitrate is  reduced to 1.1  kbps; \textbf{iii)} \textit{CQNV-v3}:  bits of LSPs are reduced to 23, and bits of pitch and energy is reduced to 12, so 1  kbps.

\begin{table}[ht]
\centering
\caption{Bit allocation for a 40 ms packet. Comparison of our three different versions and Codec2 at 1.2 kbps.}
\label{tab:bit allocation}
\setlength{\tabcolsep}{0.9mm}{%
\begin{tabular}{lllll}
\toprule[1pt]                 & Codec2 & CQNV-v1 & CQNV-v2 & CQNV-v3 \\ \hline
LSPs              & 27 bits         & \textbf{23} bits           & 27 bits           & \textbf{23} bits \\
Pitch and Energy & 16 bits         & 16 bits           & \textbf{12} bits           & \textbf{12} bits \\
Voicing level              & 4 bits          & 4 bits            & 4 bits            & 4 bits  \\
Spare            & 1 bit           & 1 bit             & 1 bit             & 1 bit  \\ \hline
Total            & 48 bits         & 44 bits           & 44 bits           & \textbf{40} bits \\  \bottomrule[1pt]
\end{tabular}%
}
\end{table}

We train HiFi-GAN using one NVIDIA RTX 3090  GPU. Our batchsize is set to 32. For the generator,  we set kernel sizes of transposed convolutions, the dilation, and kernel sizes of the residual stack to [5,4,4,2],[1,3,5],[3,7,11]. The FFT, window, and hop sizes are respectively set to 1024, 640, and 160. The optimizer, weight decay, and learning rate decay of the network are consistent with those in \cite{hifigan}. The adversarial training lasts for about 800k steps. 
\subsection{Evaluation}

\textbf{Evaluation Metrics.} The evaluation of the proposed method involves both subjective and objective metrics. To measure the subjective quality of the reconstructed speech, we utilized the MUSHRA methodology~\cite{mushra}. A group of ten listeners, including four females and six males aged between 19 to 26, participated in the subjective listening tests.  Twenty utterances, randomly selected from the test set, were evaluated. As for objective metrics, we employ the  ViSQOL~\cite{2020visqol} to measure the objective quality of the proposed method. Objective evaluation and ablation studies are measured by “speech” ViSQOL, which operates on speech resampled at 16 kHz. All the utterance signals are resampled to 16 kHz in our experiments.

\textbf{Baselines.} We compare our proposed method with both traditional codecs and neural codecs, including Codec2~\cite{jmv-codec2} at 3.2 kbps, Opus~\cite{valin2012opus} at 6 kbps, Encodec~\cite{2022encodec} at 3 kbps, and Lyra~\cite{lyra} at 3 kbps which is trained on thousands of hours of speech
data with speakers in over 70 languages\footnote{\url{https://github.com/google/lyra}}. For Encodec, we select the 24 kHz pre-trained model to synthesize speech at 3 kbps without using  Transformer language model\footnote{\url{https://github.com/facebookresearch/encodec}}. In addition, we use Speex \cite{valin2007speex} at 4 kbps as a low anchor.

\section{Results}
\subsection{Subjective test}

\begin{figure}[h]
    \centering
    \includegraphics[width=0.45\textwidth]{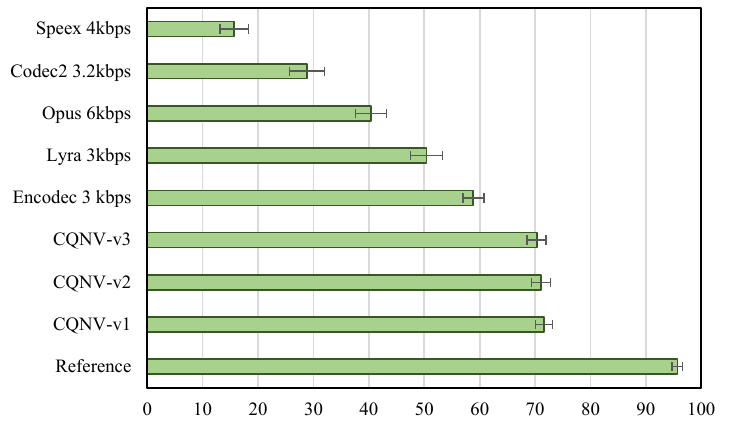}
    \caption{MUSHRA subjective test. The indicated interval in
black represents the 95\% confidence interval for each score.}
    \label{fig:mushra}
\end{figure}

Figure \ref{fig:mushra} shows the MUSHRA score as a function of codec bitrate, where
higher values and better perceptual quality. We observe
that the three versions of the proposed method outperform all
the baselines, including both traditional parametric and
neural codecs, especially showing absolute advantages compared to traditional speech codecs. The proposed method obtains obviously higher MUSHRA scores than the traditional coders, with a sharp decrease in the bitrates.
Notice CQNV at 1 kbps reaches better performance than Lyra at 3 kbps and Encodec at 3 kbps, proving our proposed method's effectiveness.

Figure \ref{fig:yuputu} illustrates a test sample's spectrograms. It can be seen from the area indicated by the dotted box that the spectrogram in the medium and low frequency band of speech synthesized by CQNV well matches the original signal. This indicates that CQNV reconstructs the harmonic structure of medium and low frequency bands better than that of Lyra and Encodec. 

\subsection{Objective test}

For objective evaluation, we use all samples from the test set. We estimate the objective quality of the speech by computing ViSQOL.  As shown in Table \ref{tab:objective}, we compare our methods with Lyra at 3 kbps and Encodec at 3 kbps. 
Table \ref{tab:objective} shows that our proposed method \textit{CQNV-v1} outperforms all other models even if they use more bandwidth. We further observe that \textit{CQNV-v1} achieves
higher ViSQOL than \textit{CQNV-v2}, indicating that there is much
redundancy in parameters LSPs than in parameters pitch and
energy in Codec2.  We also notice that the ViSQOL of \textit{CQNV-
v3} is comparable to that of \textit{CQNV-v2} with saving 4 bits, demonstrating that our method effectively removes the redundancy in speech parameters.

\begin{table}[h]
\centering
\caption{Objective evaluation of CQNV based on ViSQOL. }
\label{tab:objective}
\setlength{\tabcolsep}{7.0mm}{%
\begin{tabular}{lll}

\toprule[1pt]
Method  & Bitrate  & ViSQOL \\ \hline
CQNV-v1 & 1.1 kbps & \textbf{3.123}   \\
CQNV-v2 & 1.1 kbps & 3.030  \\
CQNV-v3 & 1 kbps & 2.957 \\
Lyra & 3 kbps & 2.865 \\
Encodec & 3 kbps & 3.083 \\

\bottomrule[1pt]
\end{tabular}%
}
\end{table}

\subsection{Ablation studies}

We conduct several additional experiments to evaluate the effectiveness of the parameters processing module and the dynamic hyper-parameter. We compare our method with the one without the processing module and another without a dynamic hyper-parameter in the same setup. As shown in Table \ref{tab:ablation}, we can see that the processing module and the hyper-parameter significantly improve the performance in speech reconstruction. We also notice a slight degradation between the non-dynamic hyper-parameter and dynamic hyper-parameter, which is more evidence of the effectiveness of this processing module.

\begin{table}[h]
\centering
\caption{Objective evaluation of CQNV-v3 at 1kbps. Ablation studies validate the effectiveness of the parameters processing module and the dynamic hyper-parameters. }
\label{tab:ablation}
\setlength{\tabcolsep}{5.0mm}{%
\begin{tabular}{ll}

\toprule[1pt]
Method    & ViSQOL \\ \hline
CQNV-v3  & \textbf{2.957} \\
~~ -v3(w/o.$\lambda_{\mathrm{FM}}$)  & 2.918 \\
~~ -v3(w/o.processing)  & 2.866   \\
~~ -v3(w/o.$\lambda_{\mathrm{FM}}$ and processing)  &2.836\\ 

\bottomrule[1pt]
\end{tabular}%
}
\end{table}

\section{Conclusions}
In this work, we present a novel framework that combines coarsely
quantized bitstream with HiFi-GAN to lower the bitrate while preserving high-quality performance. Our proposed method involves coarse LSPs, pitch and energy quantization while maintaining compatibility with existing codecs. This approach removes extra information that is not usually used by neural vocoders, and simplifies the codebook design of quantizers in traditional parametric codecs. In addition, incorporating a processing module based on traditional speech coding parameters' bitstream and introducing a dynamic hyper-parameter to balance the losses during training has been shown to enhance the quality of decoded speech. Our experimental results demonstrate the effectiveness of the proposed method in reducing the redundancy in speech parameters and improving the performance of the vocoder in speech reconstruction.
\section{Acknowledgement}
This work was supported in part by the Special Fund of Hubei Luojia Laboratory (No. 220100019), the Hubei Province Technological Innovation Major Project(No. 2021BAA034) and the Fundamental Research Funds for the Central Universities (No. 2042023kf1033).
\bibliographystyle{IEEEtran}
\bibliography{mybib}

\end{document}